# MEETING TIGHT FREQUENCY REQUIREMENT OF ROUNDED DAMPED DETUNED STRUCTURE

T. Higo, Y. Funahashi, Y. Higashi, N. Hitomi, T. Suzuki, K. Takata, T. Takatomi, N. Toge
and Y. Watanabe, KEK, Tsukuba, Ibaraki, 305-0801, Japan
Z. Li, R. H. Miller and J. W. Wang, SLAC, Stanford, CA 94305, USA

## Abstract

Following successful design and fabrication of damped detuned structures, the R&D for the accelerating structures of the NLC/JLC linear collider project proceeded to studies of Rounded Damped Detuned Structure with curved cross section of the cavity shape for increased shunt impedance. The important features of the structure are the accurately tuned accelerating mode frequency and the distribution of the first dipole modes smooth and close to the design distribution. These requirements were met based on the high-accuracy diamond turning with its capability to realize the periphery tolerance of two microns. The lowest dipole mode frequencies scattered by 0.6 MHz RMS. The error in the accelerating mode frequency averaged over a structure was 0.1 MHz by applying a feed-forward method.

## 1 INTRODUCTION

We have been developing detuned structures, DS, and damped detuned structures, DDS, for the X-band main linacs of linear colliders[1]. The essential features of those structures to realize the desired wake field are the good alignment of the constituent cells and the realization of the design frequency distribution of the dipole modes. These features have been realized basically by utilizing the present-day high-accuracy diamond turning technology[2]. The rounded damped detuned structure, RDDS, is a further modification from DDS by rounding the cell shape to get a higher shunt impedance. The basic fabrication technology is described in [3]. In the present paper, we describe how we obtained the required frequency characteristics of both the accelerating mode and the first dipole modes from the electrical point of view.

## 2 FREQUENCY REQUIREMENTS

### 2.1 Random error of accelerating mode

Random frequency errors of the accelerating mode create small reflections between the cells and small phase slips along the structure. For instance, a random error of 3 MHz RMS results in 1% loss in acceleration connected to the amplitude decrease along the structure. The same amount of error gives a phase slip which cause the acceleration loss of 1%, though it can be easily compensated by an offset in the input RF phase. In order to satisfy a criteria of less than 1% loss, we set the random error of accelerating mode at 3 MHz RMS.

### 2.2 Systematic error of accelerating mode

Even if the systematic error of 1 MHz in accelerating mode exists, the loss in the acceleration is well less than 1% if we choose some offset in feeding RF phase. However, 1 MHz error makes a phase slip of 40 degrees over a whole structure and we think it too large considering the tolerance of the RF phase error, probably less than such a value as 10 degrees or so for the BNS damping.

We set the tolerance of 5 degrees in the integrated phase advance at any point along the structure. It is to be noted that this criteria is equivalent to the frequency error of 0.1 MHz if the error is uniform over a structure.

### 2.3 Random error of first dipole mode

An important requirement on the frequency distribution is its smooth distribution close to the design distribution. The random error in the higher-mode frequencies was set at 3 MHZ RMS. This is based on the onset of the appreciable amount of the wake field degradation which is calculated by an equivalent circuit model taking the lowest two dipole modes into account. If this amount of error exists, the frequency error of a mode mainly spreading over 10 cells becomes 10 MHz which reaches the spacing of the mode frequencies and we naturally estimate that the frequency distribution deteriorates severely.

### 2.4 Systematic error of first dipole mode

Such systematic error as an offset over the whole structure does not affect the wake field at all. Higher components of the error distributions were studied and we estimated that the error should be within a few MHz[4].

## 3 BASIC FABRICATION PROCESS

Each accelerating cell is divided at the middle plane of the cell to form a disk. It is symmetric with respect to the Z-direction. All of the 3D geometries are milled to their final dimensions first and the diamond turning was performed in the last fabrication process.

The most sensitive 3D dimension on frequency is the width of the 3mm-wide slits. There are 8 slits in a disk and the frequency sensitivity is 1MHz/15μm if all of the slits have the same amount of error. We set the tolerance of the width to be ±15 μm.

Final 2D machining by a diamond tool with a precision lathe is performed with checking the outside diameter, OD, of all of the fabricated disks just after machining so that we monitor the tool radial positioning error. A correction is applied if it becomes too large approaching to 1 micron. The Z-direction reference was set by frequently self-cutting of the vacuum chuck. Every relative movement of the tool is assumed to be precise enough. By taking these processes, we believe that the accuracy of the 2D machining is better than the periphery tolerance of 2 μm, which is equivalent to confine the cut surface between two surfaces ±1 μm in and out from the design surface.

## 4 FREQUENCY CONTROL STRATEGY

The imposed periphery shape error of 2 μm will limit the possible frequency error to be maximum 3 MHz, even in the extremely unlucky error distribution within a cell both in the accelerating mode and in the first dipole mode. Therefore, the requirement on the random error is already satisfied.

The precise dimensions are specified at several points along a structure by using a high-accuracy 3D numerical electro-magnetic field solver. The dimensions in between are designed to be smoothly distributed connecting these points. The error is almost constant along a structure or at least varies very slowly because we apply any corrective action on dimensions in a smooth manner as a function of disk numbering. Since the frequency error is within 3 MHz even in the extremely big case, this procedure results in a smooth distribution of higher-mode frequency within a systematic error tolerance to the design distribution.

The most severe requirement is on the systematic error of the accelerating mode. Since the first dipole mode frequencies are designed to be distributed by varying the beam hole radius "a", the accelerating mode frequency should be kept the same by compensating with the cell radius "b".

The sensitivities of the resonant frequencies to these dimensions are listed in Table 1. As the relative movement of the tool during the cutting of a disk is very precise, we believe the errors in dimensions "a" and "b" are very close. Therefore, we estimate the error of $F_{acc}$ smaller than 1 MHz assuming the control of "a" and "b" within ±1 μm. However, this error is still too large.

Now let us discuss about three ways to correct the accelerating mode frequency as below.
- Undercut in "b" and measure frequency to feed back to "b" in the final step of machining.
- Cut to the design dimensions first and measure frequency. Then locally cut around "a" and "b".
- Cut to the design dimensions first and measure frequency. Based on the frequency information of previously machined disks, the feed forward in "b" is applied for the disks in the later fabrication.

Table 1: Typical frequency sensitivity on dimensions. The two numbers are those at input and output end.

|  | $F_{acc}$ | $F_{d1}$ |
|---|---|---|
| Units | MHz / μm | MHz / μm |
| df/db | -1.21 | -1.22~-1.32 |
| df/da | 0.65~0.45 | 0.03~-0.61 |
| df/db + df/da | -0.57~-0.76 | -1.18~-1.93 |

The first method was applied in the fabrication of DDS1,2. The whole structure was divided into 5 sections each consisting of 38 cells. As the cells in a section were fabricated with undercut in "b", its 2π/3 mode was measured. Then the "b" dimensions of the batch were cut with the same offset for all of the 38 disks. Though this method was successfully applied, we have not cited this method for the present fabrication of RDDS because this method requires a high accuracy for the final cutting, needs an additional step of measurement and introduces undesirable steps in frequencies between the batches.

The second method also needs an additional process before the final cutting but the frequency sensitivities to the local cuttings are much less than the first case. Assumed local cut shapes are a triangular cross section at the middle of a cell, called δb and a flat, constant-radius, cutting at a beam hole, called δa, both measured by the local increase of radius. The sensitivities of $F_{acc}$ to δb and δa have different signs so that it can be tuned by ±3 MHz with up to 6 μm in δb and δa. However, the derivative of $F_{d1}$ are both negative for most of the disks so that we need a negative offset in $F_{d1}$ if we need any correction. The correction of –9±3 MHz in $F_{d1}$ is realized with the same amount of corrections as above.

The last method was adopted for the present fabrication. Because the disks are symmetric in Z-direction and the dimensions vary slowly as disk number, we can pick up any three or six consecutive disks shorted at both ends to measure very accurately the 2π/3 average accelerating mode frequency. We can measure reasonably well before the fabrication of the following disks due to the rather slow production rate. Therefore, we can apply a smooth correction to "2b" for the disks in later fabrication once the integrated phase error tends to cumulates. This method does not require any additional process in machining or measurement.

## 5 RESULTS OF RDDS1

The width tolerance of 3mm-wide slit was set at ±15 μm, equivalent to ±1 MHz, so that all slit widths sat within it though they were drifting slowly as the production. However, steps appeared when the tool was exchanged. We initially decided not to compensate the effect of such a width variation by feeding back to "2b" because the associated frequency error was still smaller than the tolerable error of 3 MHz.

In turn, the precision of the machining by a diamond tool with a turning lathe was probed to be better than 1 MHz both in the monopole modes and the dipole modes. This was studied before the actual fabrication of RDDS1 disks with using the test disks of the same shape as those of RDDS1 but without 3D milled geometries[5].

In the actual fabrication of RDDS1 disks, the most relevant dimension "2b" was measured by the 3D coordinate measuring machine. In addition, the OD was measured to be within ±1 μm, though relative but precise. This indicates the errors in "2b" within ±1 μm assuming the good correlation between OD and "2b". These results proved the mechanical precision.

For the RF inspection, each disk was sandwiched by two flat plates to excite zero and π mode. Frequencies of the four modes, zero and π mode of $F_{acc}$, π mode of $F_{d1}$ and zero mode of $F_{d2}$ were measured as shown in Fig. 1. The population of the deviation of π mode $F_{d1}$ from a smooth fit curve is shown in Fig. 2. The fit was performed for the $F_{d1\pi}$ as a function of the design frequency synchronous to the beam with using a polynomial function up to the third order. The behaviours of all the other three frequencies are similarly very smooth. The deviation of any of the four frequencies is within 0.4 to 0.6 MHz RMS.

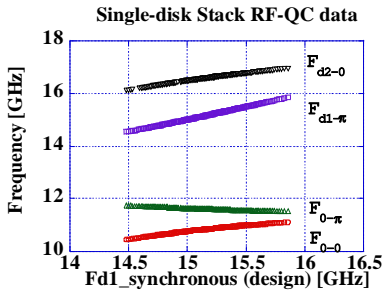

Figure 1: RF-QC results of single-disk set up.

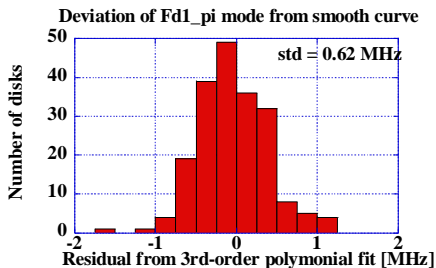

Figure 2: Each-disk frequency deviation from smooth fit.

The measured accelerating mode frequencies of the 6-disk set-ups and the associated integrated phase slip along the structure are shown in Fig. 3 with the feed-forwarded values in "2b". It is seen from the figure that the integrated phase slip was kept within 5 degrees. Since the width of 3mm slits of the disks later than #151 happened to be larger than those before, the accelerating mode frequency abruptly changed there. The resultant decrease of frequency made it necessary to apply further feed-forward values in "2b" which was gradually applied later than #160 after identifying the reason.

The frequencies of the vacuum pumping disks, which were made separately at the last stage of fabrication, were higher than the others by 0.8 MHz. The reason is not yet understood.

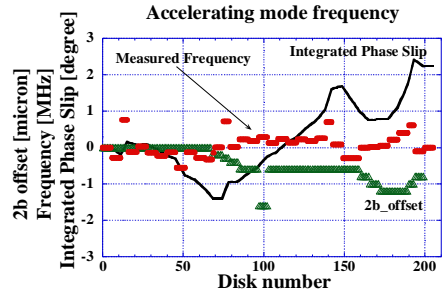

Figure 3: Accelerating mode frequency in 6-disk set up.

## 6 SUMMARY AND CONCLUSION

The random frequency errors in the relevant modes of RDDS proved to be controllable within 0.6 MHz RMS much smaller than the tolerance of 3 MHz RMS. The systematic error of the accelerating mode was controlled so as to keep the integrated phase error along the structure below the requirement at 5 degrees, equivalent to the average frequency error of 0.1MHz. This was achieved by applying the feed-forward correction on "2b". The frequency distribution of the first dipole mode was within 3 MHz to the design. From these results, we proved that the frequencies of the RDDS could be controlled well enough to meet the frequency requirements based on the present diamond turning technology with a feed-forward correction on "2b".

Even for the mass production stage, this feed-forward method can be applied but probably in a different manner. The practical method should be studied to realize a high-speed and parallel production of many structures.

## 7 ACKNOWLEGMENTS

The authors would like to thank Prof. D. Burke and Prof. M. Kihara for keeping the collaboration between two laboratories active and encouraging the authors.